\documentstyle[12pt]{article}

\newcommand{\af}{\stackrel{(f)}{a}{}}
\newcommand{\pl}{\partial_}

\newcommand{\vi}{\varphi}
\newcommand{\ol}{\overline}
\newcommand{\beq}{\begin{equation} \label}
\newcommand{\eeq}{\end{equation}}
\newcommand{\cc}{{\bf C}\ }
\newcommand{\rr}{{\bf R}}

\newcommand{\ad}{\mbox{ad }}
\newcommand{\cim}{C^\infty (M)}

\newtheorem{The}{Theorem}

\begin{document}
\large

\begin{center}
{\LARGE  Geometric quantization of generalized oscillator}
\bigskip

{\Large Sergey    V.    Zuev}\footnote{Tel./Fax  +7.8412.441604,
E-mail: szuyev@tl.ru}\\
{\small \it Department  of  Theoretical   Physics,   Penza   State
Pedagogical University,    {\rm    440039}    Penza,   Russia}
\end{center}
\bigskip

\hrule
\medskip

\noindent {\bf \normalsize Abstract}
\medskip

\noindent {\normalsize  Using  geometric  quantization procedure,  the
quantization  of  algebra  of  observables  for  physical  system  with
Ricci-flat  phase  space  is  obtained.  In  the  classical  case the
appointed physical system is reduced to harmonic oscillator  when  the
one real parameter is vanished.}\\
\hrule
\bigskip

\noindent {\bf    Keywords}:    Generalized    oscillator,   Geometric
quantization, Ricci-flat K\"ahler manifold.
\bigskip\bigskip

\noindent {\bf 1. Introduction}
\bigskip

There exist a number of procedures and methods to quantize a  physical
systems,  but it is geometric quantization that takes into account the
geometrical background (i.~e. geometry of phase space) of the physical
system. The  procedure  of  geometric  quantization  was discovered by
B.~Kostant \cite{kost1} and J.M.~Souriau \cite{sour1} in 1970.  During
last   25   years   it  was  highly  developed  by  multiply  authors.
J.M.~Tuynman in  \cite{tuy1}  has  compared  some  known  methods   of
quantization    (in    particular,   geometric   quantization)   using
2-dimensional  generalized  harmonic  oscillator,  i.~e.   hamiltonian
system with   algebra  of  observables  {\bf  su}(1,1).  For  physical
applications it would be helpful to quantize the generalized  harmonic
oscillator for  an  arbitrary  dimension.  Theorem  \ref{T-R4}  of the
present paper gives such quantization for even dimensions.

Recently a  Ricci-flat K\"ahler metric for any real dimension $4n$ was
constructed by the author \cite{myprep}.  The metric has the following
form
\beq{metr}
g_{\alpha\ol{\beta}}=u''\ol{z^\alpha}z^\beta+u'\delta^\alpha_\beta=
{a^m\over     r^2}(r^m-a^m)^{\frac{1-m}{m}}\ol{z^\alpha}z^\beta+\frac{
(r^m-a^m)^{{1\over m}}}{r}\delta^\alpha_\beta,
\eeq
where
\beq{uprim}
u'\equiv{du\over dr}=\frac{(r^m-a^m)^{1/m}}{r},
\eeq
$u''\equiv du'/dr,$    $r\equiv    \sum\limits_{\alpha=1}^m   z^\alpha
\ol{z^\alpha},$ $m\equiv 2n,$ $\rr\ni a={\rm const}.$
As it  was shown in \cite{myprep} (this proposition is almost evident)
the group of complex  isometries  of  the  metric  (\ref{metr})  is
SU($m$).  It  is notable that Ricci-flat K\"ahler phase space $M$ with
abovementioned metric has an subalgebra ${\cal F}$ of  Poisson
algebra of  $\cim$-functions  on  $M$ and ${\cal F}$ is appropriate to
quantize by geometric quantization procedure. Let us call the physical
system  with  phase  space  $M$  as  {\it generalized oscillator} and the
algebra  ${\cal  F}$  as  {\it  algebra  of  observables  of  generalized
oscillator.}
\bigskip

\noindent {\bf 2. Antiholomorphic polarization}
\bigskip

Let us  consider  a  K\"ahler polarization (see \cite{kost1,sour1} for
detailes) $F\subset TM\otimes_{\rr} \cc$ which is  determined  by  the
condition
\beq{4s2-1}
F_p=\{ X \in TM_p\otimes {\bf C}
\hskip 2pt  | X=\xi_{\alpha|p}\ad (z^{\alpha})_{p},  \xi_{\alpha|p}\in
\cc \},
\end{equation}
where $\xi_{\alpha|p}$ are $\cim$-functions of a point $p\in  M$.  The
defined polarization is called {\it antiholomorphic polarization}.
Function $f\in \cim$ preserves polarization $F$ if it  obeys  to  the
equation
\beq{4s1-11}
L_{\ad (f)} X \in F \quad\mbox{for all}\quad  X\in F.
\eeq
It is equal to the next condition
\beq{4s2-2}
[\ad (f),  \ad (z^\mu)] = \af^\mu_\beta \ad (z^\beta).
\eeq
\begin{The}{\rm \cite{amka6}}\label{4s2-th1}
Let  $M$  be a K\"ahler manifold and $F\subset TM\otimes_{\rr} \cc$
be an antiholomorphic polarization.  Function  $f\in   \cim$  preserves
polarization $F$ if and only if in every complex chart
$(U,z^\alpha,\ol{z^\alpha})$, $\alpha=1,\ldots,m$,  on $M$ the equation
\beq{4s2-8}
f=\pl\alpha \Phi \vi_\alpha (z) + \chi  (z),
\eeq
holds.  Here $\vi^\alpha (z),  \xi (z)$ are an arbitrary holomorphic
functions.
\end{The}

On the manifold $(M,g)$ where $g$ is defined by (\ref{metr})
the formula (\ref{4s2-8}) takes the following form
\beq{afun}
f=\sum_\alpha u'\ol{z^\alpha} \vi_\alpha (z) + \chi  (z),
\eeq
where $u'$ is defined by (\ref{uprim}).

Let us consider the next functions
\beq{Nfunc}
N^{\alpha\ol\beta} =           u'z^\alpha           \ol{z^\beta},\quad
\alpha,\beta=1,\ldots,m.
\eeq
It is  easy  to  show that the fuctions $N^{\alpha\ol\beta}$ preserve
polarization $F.$
From here and from theorem~\ref{4s2-th1} we find
\beq{vi}
N^{\alpha\ol{\beta}}=u'\ol{z^\sigma} \vi^{\alpha\ol{\beta}}_\sigma,
\eeq
where $\vi^{\alpha\ol{\beta}}_\sigma=z^\alpha\delta^\beta_\sigma.$

The Hamiltonian vector fields of functions $N^{\alpha\ol{\beta}}$
are defined by the following equalities
$$
V^{\alpha\ol\beta} \equiv \ad (N^{\alpha\ol\beta}) =  i\hskip 1pt
(z^\alpha \pl\beta - \ol{z^\beta} \pl{\ol\alpha}).
$$
The Poisson   brackets   of  functions  $N^{\alpha\ol\beta}$  and  the
commutators of its Hamiltonian vector fields have the next form
$$
\{ N^{\alpha\ol\beta}, N^{\mu\ol\nu}\}=
i\hskip 1pt (\delta^\beta_\mu N^{\alpha\ol\nu}-
\delta^\alpha_\nu N^{\mu\ol\beta})
$$
$$
[V^{\alpha\ol\beta}, V^{\mu\ol\nu}] =
\delta^\alpha_\nu V^{\mu\ol\beta} -
\delta^\mu_\beta V^{\alpha\ol\nu},
$$

Since $V^{\alpha\ol\beta}$ are Hamiltonian and  holomorphic  that  the
corresponding  transformations  preserve  complex structure on $M$ and
fundamental 2-form $\Omega(X,Y)\equiv g(JX,Y).$  Hence  they  preserve
metric  $g$ and they are Killing vector fields on $M$ which preserve
complex structure.  So the vector fields $V^{\alpha\ol\beta}$ form the
algebra  {\bf  su}$(m)$ of infinitesimal holomorphic isometries of the
metric (\ref{metr}).
\newpage

\noindent {\bf 3. Geometric quantization of the algebra of observables}
\bigskip

The transformations from group SU$(m)$ preserves complex structure and
metric  $g$  as well as fundamental form $\Omega.$ This means that the
action of SU($m$) on $M$ is symplectic \cite{sniat1}.  The  cohomology
group  $H^2({\bf  su}  (m),\cc)$  is  trivial  and  as it was shown in
\cite{sniat1} the action of SU($m$) on $M$ is Poisson action.

Let us consider an algebra (with respect to Poisson  brackets)  ${\cal
F}(m)$     of     linear     functions     on    $N^{\alpha\ol\beta}$,
$\alpha,\beta=1,\ldots,m.$ As it was  mentioned  in  Introduction,
this algebra is called as algebra of observables of $2m$-dimensional
generalized oscillator.  It is evident that ${\cal F} (m)$ coincides with
an  algebra  of  functions  preserving antiholomorphic polarization on
$M.$ For such a functions the following theorem exists.
\begin{The}\label{4s2-th2}{\rm \cite{amka6}}
Let $M$ be  a K\"ahler manifold, $F\subset TM\otimes_{\rr} \cc$
be an  antiholomorphic  polarization and ${\cal F}_F (M) \subset \cim$
be an algebra of functions on $M$ which preserve polarization $F$.
Then in every chart $(U,z^\alpha,\ol{z^\alpha})$,  $\alpha=1,\ldots,m$,
the quantization ${\cal Q}$ of ${\cal F}_F  (M)$  is  defined  by  the
formulae
\beq{4s2-14}
{\cal Q} (f) \psi\cdot\mu_0 = (\chi + \hbar (\vi_\sigma \pl\sigma +
\frac{1}{2} \pl\sigma \vi_\sigma )) \psi\cdot\mu_0,
\eeq
where $f=\vi_\tau \pl\tau \Phi + \chi \in {\cal F}_F$,  $\chi,\psi$ is
holomorphic  functions  on  $U$  and  $\mu_0$ is non-vanished at every
point of $U$ section of the Hermitian  vector  bundle  ${\cal  L}$
over $M.$
\end{The}

One  can  construct the quantization of algebra ${\cal  F}(m)$   using
theorem~\ref{4s2-th2}.  As far as the map ${\cal Q}$ is linear,  it is
sufficient  to  define  the  action  of  ${\cal   Q}$   on   functions
$N^{\alpha\ol\beta}$.

Using (\ref{vi}) we find from (\ref{4s2-14})
\beq{quant}
({\cal Q} (N^{\alpha\ol\beta})) \psi\cdot\mu_0 =
\hbar \left(z^\alpha \frac{\partial \psi}{\partial z^\beta} +
\frac{1}{2} \delta_\beta^\alpha \psi\right)\cdot\mu_0.
\eeq
The last formula defines quantization of algebra ${\cal  F}(m)$
of observables of $m$-dimensional generalized oscillator.

By summation (\ref{quant}) with $\alpha=\beta$ from $1$ to $m$ we have
$$
({\cal Q} (H)) \psi\cdot\mu_0 =
\hbar (z^\sigma \frac{\partial \psi}{\partial z^\sigma} +
\frac{m}{2}\psi)\cdot\mu_0,
$$
Therefore, the eigenvalues of operator ${\cal Q} H$ are defined by  the
next relation
$$
\lambda_l =  \hbar \left(l+\frac {m}{2}\right),  \qquad l=0,1,2,\ldots
\quad
$$
and coincide with energy levels of harmonic oscillator in flat space.
\bigskip

\noindent {\bf 4. Corollary}
\bigskip

The main result of the paper can be formulated as the following
\begin{The}\label{T-R4}
Let $(M,g)$  be  K\"ahler  Ricci-flat  space,  dim${}_\rr$~$M=2m=4n$  with
metric   $g,$   defined   by  {\rm  (\ref{metr})}.  Let  $F$  be  an
antiholomorphic polarization defined by {\rm (\ref{4s2-1})} and ${\cal
F}  \subset  \cim$  be an algebra of functions on $M$ which are linear
functions  on  variables  $N^{\alpha\ol{\beta}}$   defined   by   {\rm
(\ref{Nfunc})}.  Then  in  every  chart  $(U,z^\alpha,\ol{z^\alpha})$,
$\alpha=1,\ldots,m,$ the quantization ${\cal  Q}$  of  ${\cal  F}$  is
defined  dy  operators {\rm (\ref{quant})} where $\psi$ is holomorphic
function on  $U$  and  $\mu_0$  is  non-vanished  on  $U$  section  of
Hermitian vector bundle ${\cal L}$ over $M.$
\end{The}
\bigskip

\noindent {\bf Acknowledgement}
\bigskip

I would like to thank Prof.  A.V.~Aminova  and  Dr.  D.A.~Kalinin  for
their helpful discussions.

\end{document}